\documentclass[aps,pra,twocolumn,bibnotes,showpacs]{revtex4}
\usepackage{bm}
\usepackage{mathrsfs}
\usepackage{natbib}
\usepackage{amsmath,amssymb}
\usepackage[dvips]{graphicx}

\begin{document}

\title{Strong-field ionization of atoms and molecules: The two-term saddle point method }

\author{Thomas Kim Kjeldsen}
\author{Lars Bojer Madsen}
\affiliation{Department of Physics and Astronomy, University of
Aarhus, 8000 {\AA}rhus C, Denmark}

\begin{abstract}
  We derive an analytical formula for the ionization rate of neutral atoms
  and molecules in a strong monochromatic field. Our model is based on the
  strong-field approximation with transition amplitudes calculated by
  an extended saddle point method. We show that the present two-term saddle point method
  reproduces even complicated structures in angular resolved photo electron
spectra.
\end{abstract}

\maketitle

\section{Introduction}
\label{sec:introduction}

In order to describe fully the dynamics of molecules and atoms
subject to an external laser field, one must in principle solve the
time dependent Schr\"odinger equation including all degrees of
freedom. Such \textit{ab initio} solutions are, however, impossible
for any but the most simple systems, and additionally these methods
are often only available for a few specialized theoretical research
groups. Fortunately much physical insight can be achieved by simpler
models. For example, many strong-field phenomena can be successfully
interpreted if one uses the Ammosov-Delone-Krainov (ADK) tunneling
model~\cite{adk} to describe ionization.  The success and the
analytical simplicity makes the ADK model ideal for widespread use
not only for atoms but also for diatomic~\cite{Tong02} and
polyatomic~\cite{Kjeldsen05a} molecules.

Along with the ADK model, the strong-field approximation (SFA) is
one of the most widely used models to describe detachment of anions
and ionization of atoms in intense laser fields. Compared with the
ADK model, the SFA is more suited for obtaining angular and energy
resolved spectra. The two models are in fact connected since the
tunneling rate can be obtained from the SFA in the low frequency
limit~\cite{perelomov66,Gribakin97}. The initial work by
Keldysh~\cite{keldysh} concerned ionization of hydrogen. The model
was further developed by Faisal and Reiss~\cite{faisal,reiss80} and
is commonly known as the Keldysh-Faisal-Reiss (KFR) model. Later on,
the model was extended in various ways (see Ref.~\cite{Becker05} for
a recent review), e.g., to take into account
rescattering~\cite{Milosevic98,gazibegovic-busuladzic04}, long range
Coulomb potential in the final state~\cite{Duchateau02}, multiple
electrons~\cite{Becker00,Becker02} and molecular
structure~\cite{muthbohm00}. More systematically, for short-ranged
final state interactions the SFA transition amplitude is the leading
term in an exact \textit{S}-matrix series~\cite{reiss80,Becker05}.

In order to evaluate the matrix elements that enter the expression
for the ionization rate in the SFA, one may use the saddle point
method to obtain approximate closed analytical formulas.
The saddle point method can be applied in both the length- and
velocity gauge.
In the velocity gauge, the saddle point method breaks down at
intensities below $10^{13}\, \mbox{W/cm}^2$~\cite{Requate03}.
Despite its wide-spread and long-term use -- already Keldysh applied
the saddle point method in the initial work concerned with the limit
of small momenta of the outgoing electron and Ref.~\cite{popov04}
reviews other limiting formulas -- we are not aware of a similar
study of the limitation of the saddle-point method in the length
gauge. The main purpose of the present paper is to provide a
detailed discussion of the applicability of the saddle point method
in neutral atoms and to extend the theory to cover molecules. In
this effort, we identify a straightforward extension of the
conventional saddle point method. We  call the extended theory the
'two-term' saddle point method, and we show that  the present method
increases the accuracy considerably.

The paper is organized as follows. In Sec.~\ref{sec:theory} we
outline the theory. In Sec.~\ref{sec:results} we test the accuracy
of the saddle point method by presenting result on various atoms and
molecules. Section \ref{sec:conclusion} concludes.

\section{Theory}
\label{sec:theory} The saddle point method gives very accurate
results for detachment rates of negative
ions~\cite{Gribakin97,Faria02} and a saddle point formula that
covers also neutral atoms, irrespectively of the value of the momentum of
the outgoing electron, is known~\cite{Gribakin97}. The
application of the latter formula, however, was not considered until
recently~\cite{ostrovsky05a,ostrovsky05b}. The theory outlined here
follows closely the derivation of Ref.~\cite{Gribakin97}. The
differences are that (i) we take into account molecular structure,
and (ii) the previous theory only included one term in saddle-point
evaluation of a particular integral, whereas we keep two terms to
increase the accuracy and range of applicability [see
Eq.~\eqref{eqn:int_approx} below].
Equations~\eqref{eqn:dwdq}-\eqref{eqn:aqn} below summarize the basic
formulas from Ref.~\cite{Gribakin97} and are included here for
completeness.

In the single-active-electron approximation,
we consider the direct transition of the electron in an initially bound
state $\Psi_0$ to a continuum state $\Psi_{\bm q}$ due to the linearly
polarized laser field $\bm F(t) =
\bm F_0 \cos(\omega t)$ with the period $T = 2\pi/\omega$.
We quote the expression for the angular differential ionization rate
(atomic units $\hbar = m_e = |e| = 1$ with the electron charge $e=-1$ are used throughout)
\begin{equation}
  \label{eqn:dwdq}
  \frac{dW}{d\Omega} = \frac{1}{(2\pi)^2} \sum_{n = n_\textrm{min}}^{\infty}
  |A_{\bm qn}|^2 q_n,
\end{equation}
with the transition amplitude for the $n$-photon process
\begin{equation}
  \label{eqn:aqn1}
  A_{\bm qn} = \frac{1}{T}\int_0^T \langle
  \Psi_{\bm q}({\bm r},t)|\bm
F(t) \cdot \bm r|\Psi_0({\bm r},t)\rangle dt,
\end{equation}
which is to be calculated at the momentum $q_n = \sqrt{2(n \omega - E_b -
  U_p)}$, with $E_b$ the binding energy of the initial bound electron
and $U_p = F_0^2/(4\omega^2)$ the ponderomotive potential. Since the
final momentum is real, a minimum number of photons $n_\textrm{min}$
must be absorbed.
In the SFA, the interaction between the field and the electron in
the initial state $\Psi_0(\bm r,t)$ is neglected and accordingly
$\Psi_0(\bm r,t) = \Phi_0(\bm r) \exp{(iE_b t)}$, where $\Phi_0(\bm
r)$ is the stationary solution of the field-free Schr\"odinger
equation. Additionally, interactions between the residual ion and
the continuum electron are neglected in the final state which
is then described by a Volkov wave
\begin{equation}
  \label{eqn:volkov}
  \Psi_{\bm q}(\bm r,t) = \exp\left\{ i[\bm q+\bm A(t)]\cdot \bm r -
  \frac{i}{2}\int^t[\bm q + \bm A(t')]^2dt'\right\},
\end{equation}
with the vector potential $\bm A(t) = -\bm F_0/\omega \sin(\omega
t)$. We omit the lower integration limit corresponding to an
adiabatical turn-on of the field at $t\rightarrow - \infty$.

Following Ref.~\cite{Gribakin97} we write the transition amplitude
equivalently as
\begin{eqnarray}
  \nonumber
  A_{\bm qn} &=& \frac{1}{T}\int_0^T \frac{-\kappa^2-[\bm q+\bm
  A(t)]^2}{2}\tilde{\Phi}_0[\bm q + \bm A(t)]\\
  &\times& \exp\left[i\int^t \frac{[\bm q+\bm A(t')]^2 +
  \kappa^2}{2}dt'\right]dt,
  \label{eqn:aqn}
\end{eqnarray}
with $\kappa = \sqrt{2E_b}$ and $\tilde{\Phi}_0(\bm q)$ being the
Fourier transform of $\Phi_0(\bm r)$. The transition into the
continuum takes place at large distances from the ionic core, and
hence, for the initial state, it is accurate to use the asymptotic
Coulomb form which we expand in partial waves
\begin{equation}
  \label{eqn:asymptotic}
  \Phi_0(\bm r) \sim \sum_{lm} C_{lm} r^{\nu-1}e^{-\kappa r}
  Y_{lm}({\hat{\bm{r}}}),
\end{equation}
with $\nu = Z/\kappa$ and $Z$ the charge of the residual ion. Here
we assume a general non-spherically symmetric potential. For atoms,
only one term contributes and for linear molecules in the body fixed
frame, we only include $m$ states corresponding to the projection of
the angular momentum on the internuclear axis. We determine the
asymptotic expansion coefficients $C_{lm}$ by matching the
Hartree-Fock orbital of the most loosely bound electron to the form
of Eq.~\eqref{eqn:asymptotic}~\cite{kjeldsen05b}. The Fourier
transform of Eq.~\eqref{eqn:asymptotic} is
\begin{eqnarray}
  \nonumber
  \label{eqn:fourier}
  \tilde{\Phi}_0(\bm q) &=& \sum_{lm} C_{lm}
  4\pi\left(-\frac{iq}{\kappa}\right)^l\frac{\sqrt{\pi} \kappa^\nu
  \Gamma(l+\nu+2)}{2^{l+1} \Gamma(l+\frac{3}{2})(\kappa^2+q^2)^{\nu+1}}\\
  &\times&
  F\left(\frac{l-\nu}{2},\frac{l-\nu+1}{2};l+\frac{3}{2};-\frac{q^2}{\kappa^2}\right)
  Y_{lm}(\hat{\bm{q}}),
\end{eqnarray}
where $F(a,b;c;z)$ is Gauss' hypergeomtric series. We insert the Fourier
transform above in Eq.~\eqref{eqn:aqn} and obtain
\begin{eqnarray}
  \nonumber
  A_{\bm qn} &=& -\frac{1}{T}\int_0^T \sum_{lm} C_{lm}
  \left[\frac{Q(t)}{i\kappa}\right]^l\frac{2^{1-l} \pi^{3/2} \kappa^\nu
  \Gamma(l+\nu+2)}{\Gamma(l+\frac{3}{2})[Q(t)^2+\kappa^2]^{\nu}}\\
  \nonumber
&\times&  F\left(\frac{l-\nu}{2},\frac{l-\nu+1}{2};l+\frac{3}{2};-\frac{Q(t)^2}{\kappa^2}\right)
  Y_{lm}[\hat{\bm{Q}}(t)] \\
  &\times& e^{i S(t)}dt,
  \label{eqn:aqn2}
\end{eqnarray}
with kinematical momentum $\bm Q(t) = \bm q + \bm A(t)$, and action
\begin{eqnarray}
  S(t) &=& \int^t \frac{Q(t')^2 + \kappa^2}{2}dt' \\ \nonumber
  \label{eqn:action}
  &=& n \omega t + \frac{\bm q \cdot \bm F_0}{\omega^2}\cos(\omega t) -
  \frac{U_p}{2\omega}\sin(2\omega t).
\end{eqnarray}
In a multiphoton process $n\gg1$ and the exponential factor $e^{iS(t)}$
therefore oscillates rapidly on the interval $0\le t \le T$. This fact makes
the integral difficult to evaluate directly from Eq.~\eqref{eqn:aqn2}.

\begin{figure}
  \begin{center}
    \includegraphics[width=\columnwidth]{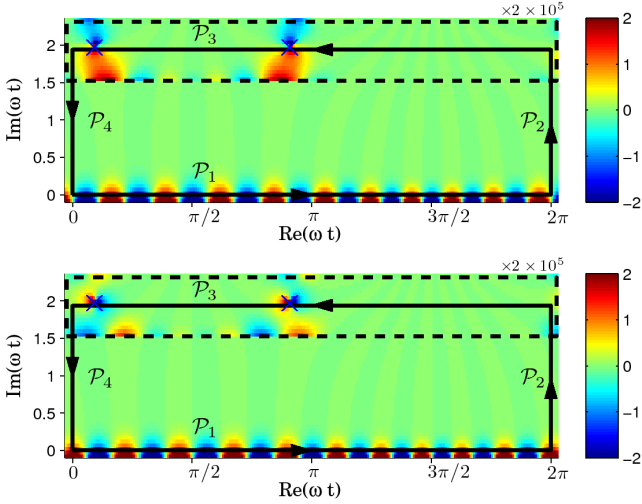}
  \end{center}
  \caption{(Color online) Integration contour.
  Upper (lower) panel: real (imaginary) part of the oscillating factor
  $e^{iS(t)}$. In the regions bounded by dashed squares,
  function values are multiplied by $2\times 10^5$ ($\textrm{Im}(\omega t)>1.5$).
  The parameters correspond to ionization of hydrogen in the polarization direction by 10 photons
  at $800\, \mbox{nm}$ at an intensity of $1\times 10^{13}\, \mbox{W/cm}^2$.
  }
  \label{fig:path}
\end{figure}

The time integral of Eq.~\eqref{eqn:aqn2} follows obviously the real
$t$ axis. From Eq.~\eqref{eqn:aqn2} and the convergence of the
hypergeomtric function, we see that the integrand is an analytical
function of $t$ except at complex instants of time satisfying
\begin{equation}
  \label{eqn:saddlepts}
  [\bm q + \bm A(t)]^2 = Q(t)^2 = - \kappa^2,
\end{equation}
where $\tilde{\Phi}_0[\bm q + \bm A(t)]$ is singular. These points coincide
with the saddle points $S'(t) = 0$ in the factor $e^{iS(t)}$. We discuss the
complex momentum in appendix~\ref{sec:momentum} and refer to
Refs.~\cite{gazibegovic-busuladzic04,Gribakin97} for the analytical
properties and evaluation of the spherical harmonics on a complex vector.
The continuation to the complex $t$ plane is straightforward as long as we
remember to treat the singularities with care. Note that the singularities
vanish when $\nu = Z = 0$ corresponding to detachment of negative ions.  In
Fig.~\ref{fig:path} we show the integrand of Eq.~\eqref{eqn:aqn2} for a
typical set of laser parameters applied to the ground state of hydrogen
($l=m=0, C_{00}=2$). The integral along the closed contour shown in
Fig.~\ref{fig:path} is zero according to Cauchy's theorem since we carefully
avoid to enclose the singularities marked by crosses. The integrand is
clearly invariant under the periodic translation ${\rm Re}(t)+i{\rm Im}(t)
\rightarrow {\rm Re}(t+T)+i{\rm Im}(t)$. Hence, the contributions to the
integral from the vertical paths $\mathcal{P}_2$ and $\mathcal{P}_4$ cancel
exactly and, consequently, the contributions along the horizontal paths must
also cancel. We can therefore equally well evaluate the integral
Eq.~\eqref{eqn:aqn2} along the negative $\mathcal{P}_3$ path.
At the expense of introducing complex and unphysical times, we see from
Fig.~\ref{fig:path} that we can calculate the integral more efficiently in
the complex $t$ plane. It is apparently much easier to
evaluate the integral along the path $-\mathcal{P}_3$ than along the real
axis $\mathcal{P}_1$. Along the former path, the factor $e^{iS(t)}$ is
nearly zero everywhere except at the two saddle points while the
same factor oscillates along the entire path $\mathcal{P}_1$. This fact is
of course the basis for any asymptotic expansion.
Strictly speaking, the saddle point method requires that we deformate the
path to pass across the saddle point in the direction of the steepest descent.
In practice, the required deformation from the horizontal line is small and has
negligible influence on the final results.

In the saddle point method outlined in Ref.~\cite{Gribakin97}, one
neglects the variations of $\bm Q(t)$ over the range where the
factor $[Q(t)^2+\kappa^2]^{-\nu}e^{iS(t)}$ has a significant
amplitude, i.e., according to Eq.~\eqref{eqn:saddlepts} we let $Q(t)
= \pm i \kappa$ in the remaining factors and obtain
\begin{eqnarray}
  \label{eqn:aqn_saddle}
  \nonumber
  A_{\bm qn} &\approx& -\sum_{lm}C_{lm}
  \Gamma(\nu+1)\left(\frac{\kappa}{\omega}\right)^\nu \sum_{\mu=1,2}
  (\pm 1)^l Y_{lm}(\hat{\bm q}_\mu) \\
  &\times&\int_{-\mathcal{P}_{3}^\mu}
  \frac{e^{iS(\phi)}}{[S'(\phi)]^\nu}d\phi,
\end{eqnarray}
where $\phi = \omega t$ and $S'(\phi) =
[Q(t)^2+\kappa^2]/(2\omega)$. We expect such an approximation to be
most accurate for $l=0$ since the factors $Q(t)^lY_{lm}[\hat{\bm
Q}(t)]$ in Eq.~\eqref{eqn:aqn2} are constant in this case. The
integrals are to be evaluated along the negative direction of the
path $\mathcal{P}_3$ near the $\mu$'th point of stationary phase. In
Ref.~\cite{Gribakin97}, the denominator was expanded as
\begin{equation}
  [S'(\phi)]^{-\nu} \approx [S''(\phi_\mu)]^{-\nu}(\phi-\phi_\mu)^{-\nu},
\end{equation}
near the saddle points. As we show in Sec.~\ref{sec:hydrogen} below,
we obtain higher accuracy if we expand $S'(\phi)$ to second order around the
saddle point
\begin{equation}
  S'(\phi) \approx
  S''(\phi_\mu)(\phi-\phi_\mu)\left[1+\frac{S^{(3)}(\phi_\mu)}{2 S''(\phi_\mu)}
  (\phi-\phi_\mu)\right],
\end{equation}
and by the first order binomial series
\begin{eqnarray}
  \nonumber
  [S'(\phi)]^{-\nu} &\approx& [S''(\phi_\mu)]^{-\nu}(\phi-\phi_\mu)^{-\nu} \\
  &&- \nu \frac{S^{(3)}(\phi_\mu)}{2 [S''(\phi_\mu)]^{\nu+1}}
  (\phi-\phi_\mu)^{-\nu+1}.
  \label{eqn:denominator}
\end{eqnarray}
With this expansion inserted in Eq.~\eqref{eqn:aqn_saddle}, the integral now
becomes a sum of two terms
\begin{equation}
  \int_{-\mathcal{P}_{3}^\mu}
  \frac{e^{iS(\phi)}d\phi }{[S'(\phi)]^\nu}\approx
  \mathcal{I}_{0\mu}(1+\mathcal{C_\mu}),
  \label{eqn:int_approx}
\end{equation}
with the conventional saddle-point term
\begin{eqnarray}
  \nonumber
  \mathcal{I}_{0\mu} &=& [S''(\phi_\mu)]^{-\nu}\int_{-\mathcal{P}_{3}^\mu}
  \frac{e^{iS(\phi)}}{(\phi-\phi_\mu)^\nu}d\phi\\
  &\approx&  \frac{i^\nu \Gamma(\frac{\nu}{2})}{2\Gamma(\nu)}
  \left[\frac{-2i}{S''(\phi_\mu)}\right]^{\frac{\nu}{2}}
  \left[\frac{2\pi i}{S''(\phi_\mu)}\right]^\frac{1}{2}
  e^{iS(\phi_\mu)}
  \label{eqn:I0}
\end{eqnarray}
and the present correction term
\begin{eqnarray}
  \nonumber
  \mathcal{C}_\mu &=& -\frac{1}{\mathcal{I}_0}\frac{\nu S^{(3)}(\phi_\mu)}{2 [S''(\phi_\mu)]^{\nu+1}}
  \int_{-\mathcal{P}_{3}^\mu}
  \frac{e^{iS(\phi)}}{(\phi-\phi_\mu)^{\nu-1}}d\phi\\
  &\approx& \frac{\nu S^{(3)}(\phi_\mu)}{(2i)^{1/2}S''(\phi_\mu)^{3/2}}
  \frac{\Gamma\left(\frac{\nu+1}{2}\right)}{\Gamma\left(\frac{\nu}{2}\right)}.
  \label{eqn:C}
\end{eqnarray}
In Eqs.~\eqref{eqn:I0} and \eqref{eqn:C}, we extended the integration limits to
infinity and used the asymptotic approximation~\cite{Gribakin97}
\begin{equation}
  \int \frac{e^{iS(\phi)}d\phi }{(\phi-\phi_\mu)^\nu}\approx
  \frac{i^\nu \Gamma(\frac{\nu}{2})}{2\Gamma(\nu)}\left[\frac{2\pi
  i}{S''(\phi_\mu)}\right]^{\frac{1}{2}}
  [-2iS''(\phi_\mu)]^{\frac{\nu}{2}} e^{iS(\phi_\mu)}.
\end{equation}
$\mathcal{I}_{0\mu}$ recovers the result of Ref.~\cite{Gribakin97} while
$\mathcal{C_\mu}$ is a correction that arizes from the higher order
expansion of $[S'(\phi)]^{-\nu}$. We present the main formula of the
present work in the next equation
\begin{widetext}
\begin{eqnarray}
\label{eqn:A-main} A_{\bm qn} \approx -\sum_{lm}C_{lm}
  \Gamma(\nu+1)\left(\frac{\kappa}{\omega}\right)^\nu \sum_{\mu=1,2}
  (\pm 1)^l Y_{lm}(\hat{\bm q}_\mu) \times \left\{ \begin{array}{ll}
  {\cal I}_{0\mu}(1+{\cal C}_\mu) & \mbox{two-term} \\
{\cal I}_{0\mu} & \mbox{one-term}
\end{array} \right..
\end{eqnarray}
\end{widetext}
With the inclusion of ${\cal I}_{0\mu}(1+{\cal C}_\mu)$,  we refer
to Eq.~\eqref{eqn:A-main} as the {\it two-term saddle point
approximation} while neglecting $\mathcal{C}_\mu$ and maintaining
only ${\cal I}_{0\mu}$ in \eqref{eqn:A-main} is referred to as the {\it
one-term saddle point approximation}.
We note from Eq.~\eqref{eqn:C} that there is no significant 
additional numerical complications involved with the inclusion of 
the second term.

We find $\phi_\mu$ from the saddle point conditions of Eq.~\eqref{eqn:action}
\begin{equation}
  \label{eqn:polyroot}
  S'(\phi_\mu) = n-z + \xi \sin \phi_\mu +2z\sin^2 \phi_\mu = 0,
\end{equation}
with $\xi = -\bm F_0 \cdot \bm q/\omega^2$ and $z=U_p/\omega$. The solutions
to Eq.~\eqref{eqn:polyroot} are
\begin{equation}
  \label{eqn:smu}
  \sin \phi_\mu = \frac{-\xi\pm i\sqrt{8z(n-z)-\xi^2}}{4z},
\end{equation}
from which it follows that
\begin{eqnarray}
  \label{eqn:cmu}
  \cos \phi_\mu  &=& \pm \sqrt{1-\sin^2\phi_\mu} \\
  \label{eqn:expsmu}
  e^{iS(\phi_\mu)} &=& (\cos \phi_\mu+i\sin\phi_\mu)^n e^{-i\cos
  \phi_\mu(\xi+z \sin\phi_\mu)} \\
  \label{eqn:spp}
  S''(\phi_\mu) &= & \cos\phi_\mu(\xi+4z \sin\phi_\mu )\\
  \label{eqn:sppp}
  S^{(3)}(\phi_\mu) &=& -\xi \sin\phi_\mu + 4z[2 \cos^2\phi_\mu-1],
\end{eqnarray}
where the signs correspond to $Q(t_\mu) = \mp i \kappa$ at the saddle
points. We combine Eqs.~\eqref{eqn:aqn_saddle},
\eqref{eqn:int_approx}-\eqref{eqn:C} and
\eqref{eqn:smu}-\eqref{eqn:sppp} to obtain the analytical approximation to
the transition amplitude, Eq.~\eqref{eqn:aqn2}.

The inclusion of nuclear motion in the molecular case was discussed
in detail for ionization~\cite{kjeldsen05b} and for high harmonic
generation~\cite{Madsen06a} within the SFA. The form of the
amplitude \eqref{eqn:aqn2} stays the same and the formulas are
straightforwardly generalized. When it comes to the assessment of
the accuracy of the saddle-point method which is the main objective
of the present paper, nuclear motion is unimportant and is left out
for clarity.

\section{Results}
\label{sec:results}

\subsection{Test case: atomic hydrogen}
\label{sec:hydrogen} First we consider ionization of ground state
hydrogen. We use this system to benchmark the accuracy of the saddle
point method against numerical integration. The atomic structure
parameters are $C_{00} = 2$, $\kappa = 1$ and $\nu = 1$.
Furthermore, the asymptotic form of Eq.~\eqref{eqn:asymptotic} is
identical to the exact wave function at all distances and the
Fourier transform is
\begin{equation}
  \tilde{\Phi}_0(\bm q) = \frac{16 \pi}{(1+q^2)^2} \frac{1}{\sqrt{4\pi}}.
\end{equation}
The spherical harmonic and the hypergeomtric function in
Eq.~\eqref{eqn:aqn2} are both constant and it is therefore exact to neglect
variations therein around the saddle point when we derive
Eq.~\eqref{eqn:aqn_saddle}.

After choosing the alternative integration path $-\mathcal{P}_3$ of
Fig.~\ref{fig:path}, the transition amplitude reduces to
\begin{eqnarray}
\label{eqn:A-brint}
  A_{\bm qn} = -\frac{8\pi}{\sqrt{4\pi} \omega}
  \frac{1}{2\pi} \times \sum_{\mu=1,2} \left\{ \begin{array}{ll}
  {\cal I}_{0\mu}(1+{\cal C}_\mu) & \mbox{two-term} \\
{\cal I}_{0\mu} & \mbox{one-term}
\end{array} \right. 
\end{eqnarray}
Here we test the accuracy of Eq.~\eqref{eqn:A-main} [or
\eqref{eqn:A-brint}], and in particular the difference between the
one-term (only ${\cal I}_{0\mu}$ included) and the two-term (${\cal
I}_{0\mu}(1+{\cal C}_\mu)$ included) saddle point formulas.
\begin{figure}
  \begin{center}
    \includegraphics[width=\columnwidth]{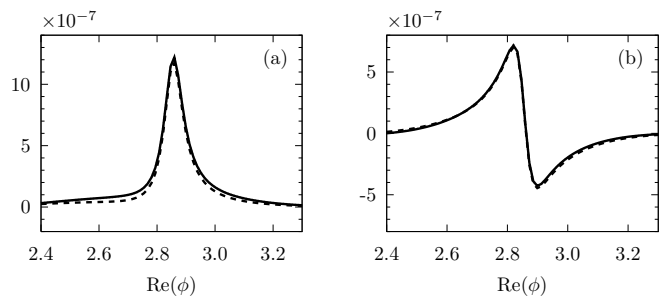}
  \end{center}
  \caption{(a) Real and (b) imaginary parts of the integrand
  $e^{iS(\phi)}[S'(\phi)]^{-\nu}$ along the integration path $-\mathcal{P}_3$ in the neighborhood
  of the second saddle-point in Fig.~\ref{fig:path}, solid. The long dashed line is the
  approximation around the second saddle point including
  only ${\cal I}_{0\mu}$.
  The two-term approximation including ${\cal I}_{0\mu}(1+{\cal C}_\mu)$ is the short dashed line which
  overlaps completely the solid line.
  The parameters of the laser are as in Fig.~\ref{fig:path}.}
  \label{fig:integrand}
\end{figure}
In Fig.~\ref{fig:integrand} we show the integrand along the
integration path in the neighborhood of the second saddle point
(see Fig~\ref{fig:path}). Additionally, we show the results of the
approximations using the one- and two-term expansion around the
second saddle point in Eq.~\eqref{eqn:denominator}. We see that the
one-term expansion recovers quite well the peak structure around the
saddle point. In the wings of the peak, the two-term expansion is
significantly better, which is most evident from the real part of
the integrand, panel~(a). We have integrated the integral
numerically and obtained the value $5.64\times 10^{-8}$ while we
obtain the values $4.55\times 10^{-8}$ and $5.39\times 10^{-8}$ for
the one- and two-term saddle point approximation,
Eq.~\eqref{eqn:int_approx} summed over the two saddle points.

\begin{figure}
  \begin{center}
    \includegraphics[width=\columnwidth]{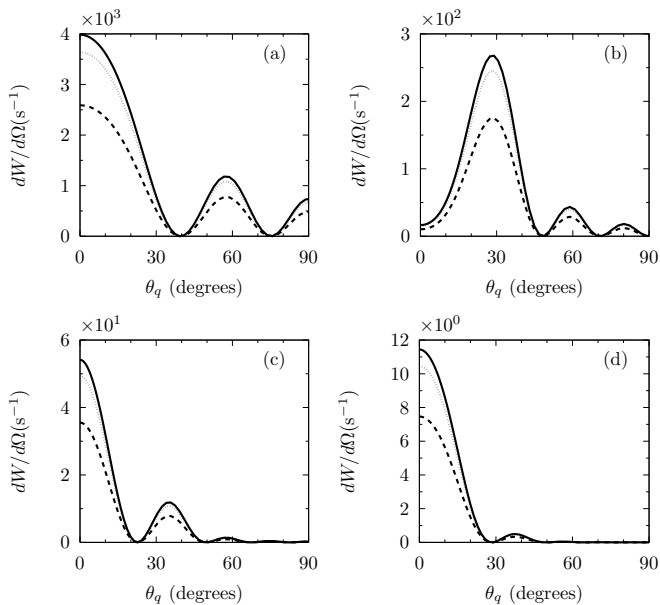}
  \end{center}
  \caption{(a)-(d) Angular differential ionization rate of hydrogen
  for the lowest number of
  photon absorptions, $n = 10-13$ ordered by increasing $n$.  The
  solid curve is obtained by numerical integration while long and short
  dashed curves are obtained by the saddle point method with the
  one- and two-term approximation, respectively. The
  laser wavelength is $800\, \mbox{nm}$ and the intensity is $1 \times
  10^{13}\, \mbox{W/cm}^2$.}
  \label{fig:h-angular}
\end{figure}
Having seen that the saddle point method is accurate in the single
case above, namely ionization parallel to the field by 10 photons,
we now present the $n$-photon angular differential rates at varying
photon orders in Fig.~\ref{fig:h-angular}. Here $\theta_q$ is the
polar angle of the outgoing electron with respect to the
polarization axis. Figure \ref{fig:h-angular} shows that both saddle
point methods predict an angular structure in close agreement with
the numerical integration. The rates obtained by the single-term
approximation are, however, around $35\%$ too small. The two-term
approximation is significantly better with an accuracy within
$10\%$.

\begin{figure}
  \begin{center}
    \includegraphics[width=\columnwidth]{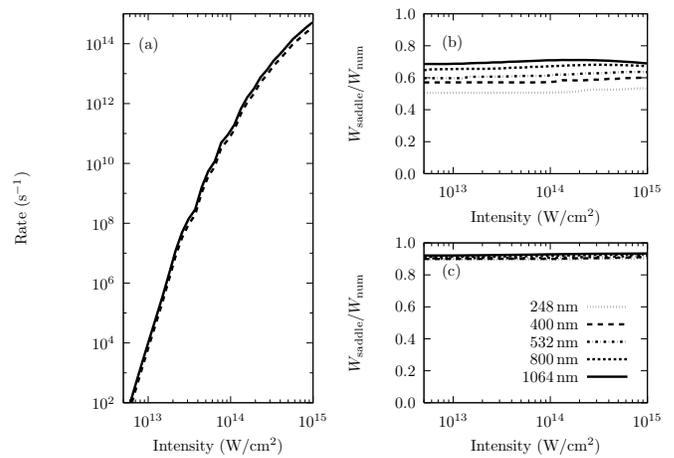}
  \end{center}
  \caption{
  (a) Absolute
  rates at a wavelength of $800\, \mbox{nm}$. The result of
  numerical integration is indicated by the solid line while the long- and
  short dashed lines are the one- and two-term saddle point approximation,
  respectively.
  (b) and (c) Ratio between total ionization rates obtained by the saddle point
  method and numerical integration for varying wavelength and intensity.
  We show saddle point results for the one- and two-term approximation
  in panels (b) and (c), respectively.}
  \label{fig:h-ratio}
\end{figure}

In our final test, we consider the total ionization rate integrated
over all angles of the outgoing electron. In
Fig.~\ref{fig:h-ratio}~(a) we present total ionization rates at
$800\, \mbox{nm}$ obtained with the numerical integration and the
one- and two-term saddle point approximation. All three methods
produce results in quite good agreement over many orders of
magnitude on the scale shown in the figure. In order to investigate
the accuracy of the saddle point method in some more detail, we
calculate the ratio $W_\textrm{saddle}/W_\textrm{num}$ between the
rates obtained by the saddle point method and the numerical
integration. Figures \ref{fig:h-ratio}~(b) and (c) present the
results for various wavelengths and intensities.
Again, we use both the one- and two-term approximation, i.e., we study
the results of Eq.~\eqref{eqn:A-main} with ${\cal I}_{0\mu}$ and
${\cal I}_{0\mu}(1+{\cal C}_\mu)$, respectively. First, we see that
the accuracies of both saddle point methods are nearly independent
of the intensity for each fixed value of the wavelength. From
panel~(b) we note that the results of the one-term approximation
depend significantly on the wavelength. The error is up to a factor
of two for the shortest wavelength $248\, \mbox{nm}$ while the error
decreases with increasing wavelength. The two-term approximation, on
the other hand, produces much more accurate results, panel~(c). The
rates are approximately $10\%$ too small for all intensities and
wavelengths considered. Even though the simple single-term saddle
point approximation is somewhat poorer than the two-term
approximation the error in the hydrogenic case is approximately
constant over twelve orders of magnitude and is not expected to be
of major significance compared with the approximations in the SFA
itself. In the rest of the paper, we use only the two-term saddle
point formula. 

\subsection{Ionization of atoms}
\label{sec:atoms}

In this section we show results for the noble gas atoms where the
active electron initially occupies an orbital in the filled
\textit{p} shell. We calculate the rates for each of the states
$m=-1,0,1$ and multiply the result by two corresponding to two
equivalent electrons in each orbital. We take the atomic structure
parameters $C_{lm}$ and $E_b$ from Ref.~\cite{kjeldsen05b}.

\begin{figure}
  \begin{center}
    \includegraphics[width=\columnwidth]{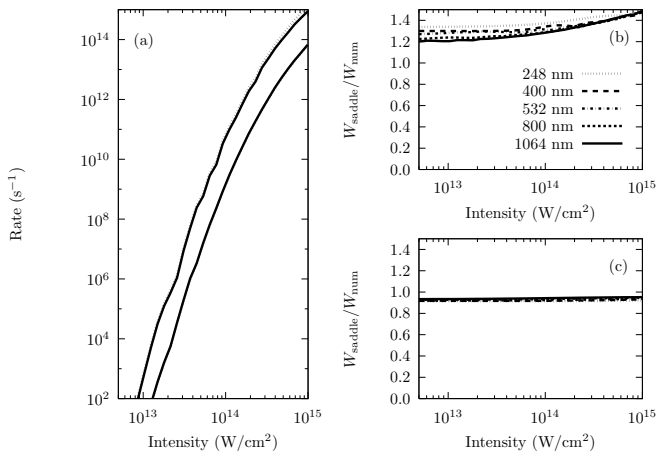}
  \end{center}
  \caption{
  (a) Absolute rates for argon at a wavelength of $800\, \mbox{nm}$. The results of
  numerical integration are indicated by the solid lines while the
  short dashed lines are the two-term saddle point approximation. The upper and lower
  sets of curves are for $m=0$ and $m=1$, respectively.
  (b) Ratio between total ionization rates from the $m=0$ state
  obtained by the saddle point
  method and numerical integration for varying wavelength and intensity.
  (c) Similar to (b) for the $m=1$ state.
  }
  \label{fig:ar-rates}
\end{figure}
In Fig.~\ref{fig:ar-rates} we present the absolute and relative rates for
the argon atom.
As in the case of hydrogen, the saddle point method is accurate over
many orders of magnitude, Fig.~\ref{fig:ar-rates}~(a).
Interestingly, the saddle point method is slightly better for the
$m=1$ [panel (c)] state compared with the $m=0$ state [panel (b)].
This $m$-dependent accuracy turns out to be important in the
molecular case as we show in Sec.~\ref{sec:molecules} below.

\begin{figure}
  \begin{center}
    \includegraphics[width=\columnwidth]{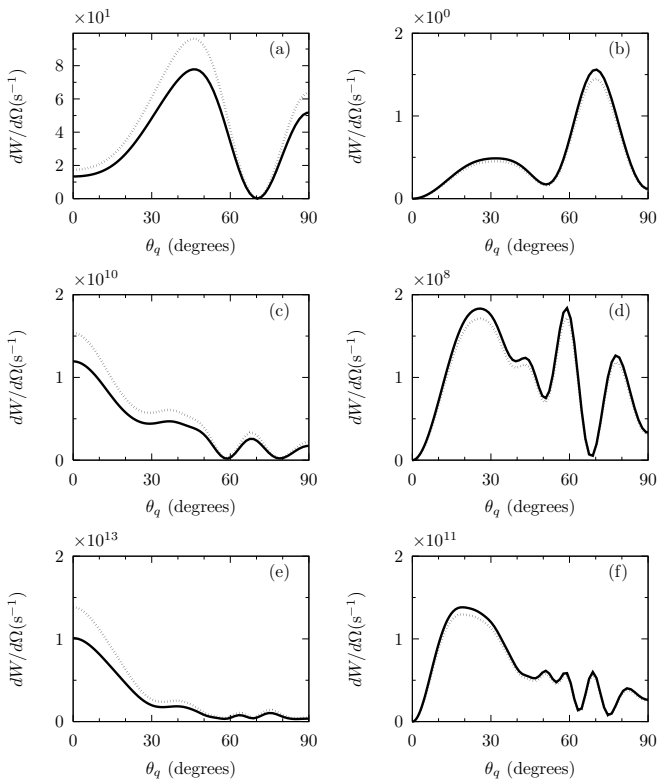}
  \end{center}
  \caption{Left (right) column: Angular differential rate for argon
  in the $m=0$ ($m=1$) state. Solid lines are obtained by numerical
  integration while the dashed lines are by the two-term saddle point method.
  The laser wavelength is $800\, \mbox{nm}$ and the
  intensities are (a)-(b) $1\times 10^{13}\, \mbox{W/cm}^2$,
  (c)-(d) $1\times 10^{14}\, \mbox{W/cm}^2$ and (e)-(f)
  $3\times 10^{14}\, \mbox{W/cm}^2$}
  \label{fig:ar-angular}
\end{figure}
Figure~\ref{fig:ar-angular} shows angular differential rates summed over all
photon absorptions at a wavelength of $800\, \mbox{nm}$ for different intensities.
We show the results for both the $m=0$ and $m=1$ state and again we see that
the saddle point method works better for the $m=1$ state. The general
features in the angular spectra are, however, very well reproduced in all
cases.

We mention in closing that the results for krypton and xenon are
very similar to argon and are therefore omitted here for brevity.

\subsection{Ionization of molecules}
\label{sec:molecules}

In the molecular case, the calculations are most conveniently
performed in the laboratory fixed frame with the $z$ axis parallel
to the laser polarization. Accordingly, we must express the initial
wave function in this frame. The wave function and asymptotic
expansion coefficients $C_{lm}$ are, however, most naturally
expressed in the body-fixed molecular frame. If the body-fixed frame
is rotated by the Euler angles $(\alpha,\beta,\gamma)$ with respect
to the laser polarization, we rotate the wave function into the
laboratory fixed frame by the rotation operator $\Phi_0(\bm r)
\rightarrow D(\alpha,\beta,\gamma)\Phi_0(\bm r)$. The rotation
operation effectively allows us to express the asymptotic
coefficients in the laboratory frame (LF) by the corresponding
coefficients in the molecular frame (MF)
\begin{equation}
  C_{lm}^\text{LF} = \sum_{m'=-l}^l
  \mathscr{D}^{(l)}_{mm'}(\alpha,\beta,\gamma)C_{lm'}^\text{MF},
  \label{eqn:clm_rot}
\end{equation}
where $\mathscr{D}^{(l)}_{m'm}(\alpha,\beta,\gamma)$ is a Wigner
rotation function~\cite{zare,brink}. For linear polarization and the
linear molecules considered in the present work, we only need to
consider rotation by the angle $\beta$ between the molecular and
field axes. We refer to Ref.~\cite{kjeldsen05b} for the coefficients
$C_{lm}^\text{MF}$.

\begin{figure}
  \begin{center}
    \includegraphics[width=0.6\columnwidth]{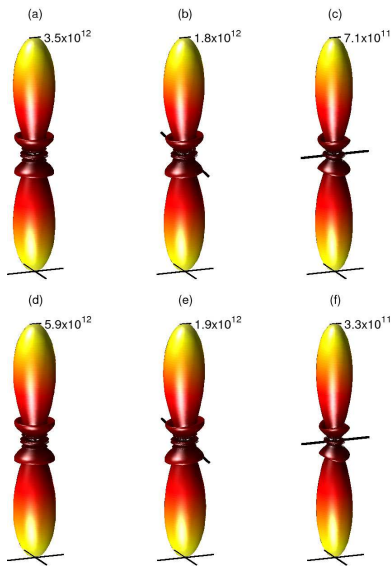}
  \end{center}
  \caption{(Color online) Angular differential ionization rate for N$_2$ aligned at an
  angle of $0^\circ$ [(a) and (d)], $45^\circ$ [(b) and (e)], and $90^\circ$
  [(c) and (f)] with respect to the polarization. Panels (a)-(c) are
  obtained by numerical integration and panels (d)-(f) by the two-term saddle point
  method.  In all panels, the polarization direction is vertical. In panels (a) and (d) the molecular
  axis is along the vertical polarization. In the other panels the
  orientation of the
  molecular axis is indicated by the line through the origin.  All three
  coordinate axes are scaled equally and the rates are given in units  of
  $\mbox{s}^{-1}$ on the scale indicated in each panel.  The laser wavelength is
  $800\, \mbox{nm}$ and the intensity $2\times 10^{14}\, \mbox{W/cm}^2$. }
  \label{fig:n2-angular}
\end{figure}
Figure~\ref{fig:n2-angular} presents angular differential rates for
differently aligned N$_2$ molecules which ionize from the doubly
occupied $3\sigma_g$ orbital. We show both numerical [(a)-(c)] and
two-term saddle point results [(d)-(f)]. First we note that the two
methods agree perfectly on the shape of the angular distribution for
all alignment angles, $\beta$. The structures are also in good
agreement with Ref.~\cite{kjeldsen04a}, where we used an atomic
centred Gaussian basis expansion for the initial state and
calculated the transition amplitude numerically. Secondly, the
overall structure is nearly independent of $\beta$. The angular rate
is simply much favoured along the polarization direction in all
geometries. This observation agrees well with the predictions of
tunneling theory where the electron by assumption escapes along the
polarization axis (For Keldysh parameter $\gamma =\kappa \omega/F_0
\ll 1$, the ionization dynamics is tunneling like. In
Fig.~\ref{fig:n2-angular}, $\gamma = 0.81$, i.e., approaching the
tunneling regime.)

\begin{figure}
  \begin{center}
    \includegraphics[width=0.6\columnwidth]{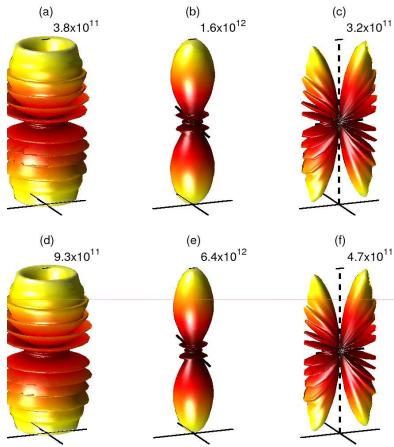}
  \end{center}
  \caption{(Color online) Similar to Fig.~\ref{fig:n2-angular} but for O$_2$.}
  \label{fig:o2-angular}
\end{figure}
Figure~\ref{fig:o2-angular} shows the angular differential rate for
O$_2$ which ionizes from the two half-filled degenerate  $\pi_g$
orbitals, one with $m=1$ and one with $m=-1$. We show the results
for a single electron with projection $m=1$ and note that the rate
is similar for $m=-1$. Again we see that the two methods predict the
exact same complex angular structures. The structures can be
understood from the symmetry of the initial wave function. The
initial $\pi_g$ orbital has zero amplitude along- and in the plane
perpendicular to the molecular axis and the nodal structure of the
wave function forbids the electron to be emitted along the vertical
polarization axis when this axis coincides with a nodal
plane~\cite{Kjeldsen05a}.

\begin{figure}
  \begin{center}
    \includegraphics[width=\columnwidth]{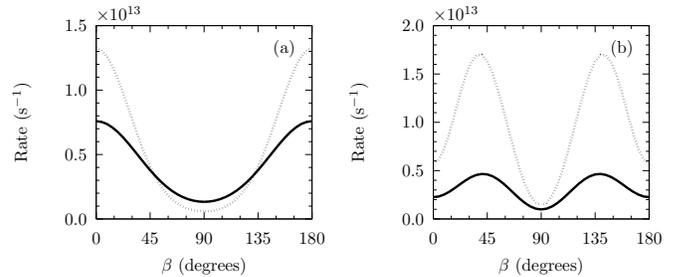}
  \end{center}
  \caption{Alignment dependent ionization rates for (a) N$_2$ and (b) O$_2$.
  The solid lines are obtained by numerical integration and the dashed line
  by the two-term saddle point method.
  The parameters of the laser are as in Fig.~\ref{fig:n2-angular}.}
  \label{fig:n2-o2-beta}
\end{figure}

In Figs.~\ref{fig:n2-angular} and \ref{fig:o2-angular} we see that
even though the angular structures agree perfectly, the absolute
scales differ by up to a factor of four. We therefore finally turn
to a discussion of the alignment dependent rate. We calculate the
total rate integrated over all angles of the outgoing electron and
show the results in Fig.~\ref{fig:n2-o2-beta}.
The figure shows that both methods agree that the rate for N$_2$ is
maximized when the molecule is aligned parallel to the polarization
axis ($\beta=0$) and minimized when aligned perpendicularly ($\beta
= 90^\circ$). Such an alignment dependence is also seen
experimentally~\cite{corkum03}. For O$_2$ both methods also agree
that the rate is maximized around an alignment angle of $40^\circ$.
For both molecules, however, the two-term saddle point method
predicts a too large variation compared with the numerical
integration. The reason for this disagreement lies in the
$m$-dependent accuracy of the saddle point method as we discussed in
Sec.~\ref{sec:atoms}. When we rotate a molecule we change the
expansion of the initial wave function in the laboratory fixed
spherical harmonics. The rotation operation mixes the different
$m$-states according to Eq.~\eqref{eqn:clm_rot}. Since the part of
the transition amplitude that corresponds to each of the $m$-states
can be either slightly too large or too small by the saddle point
method, the overall accuracy depends on the partial wave
decomposition after the rotation. For the atomic $l=1$ and $m=0,\pm
1$ states of Sec~\ref{sec:atoms}, the two-term saddle point method
is still quite accurate and the small differences reported there
compared with the numerical integration  cannot account for the
disagreement between the two methods in Fig.~\ref{fig:n2-o2-beta}.
For the molecules, however, we have included angular momentum states
up to $l=4$ and it turns out that the saddle point method becomes
increasingly inaccurate with increasing $l$. If the active electron
initially occupies an orbital with a component of non-zero angular
momentum, we expect the saddle point method to be somewhat poorer
than for $l=0$ as we discussed in deriving
Eq.~\eqref{eqn:aqn_saddle}. In Eq.~\eqref{eqn:aqn_saddle}, we
evaluate the factors $Q(t)^lY_{lm}[\hat{\bm Q}(t)]$ and the
hypergeomtric function at the saddle points. This approximation is
naturally most accurate if $\bm Q(t)$ is nearly constant in the
vicinity of the saddle points. In appendix~\ref{sec:momentum},  we
calculate $Q(t)$ along the integration contour and we see from
Fig.~\ref{fig:qt} that the variation in $\textrm{Im}[Q(t)]$ is in
fact close to maximal at the saddle points.  If we require higher
accuracy of the saddle point method, we must take at least the first
order variation in $\bm Q(t)$ into account and modify
Eq.~\eqref{eqn:aqn_saddle} accordingly.

\section{Summary and Conclusion}
\label{sec:conclusion}

Based on the length gauge SFA, we proposed a two-term saddle point
formula which is applicable to neutral atoms and molecules. We
presented calculations on various atoms and molecules with the
primary aim to test the accuracy of the method.  The two-term
saddle-point evaluation is very accurate in the case of ionization
of hydrogen while the accuracy is within $\approx 10\%$ for noble
gas atoms which undergo ionization from a \textit{p} shell.
Remarkably, the structures of the angular photo electron spectrum
predicted for all systems are in perfect agreement with numerical
calculations. We have identified that the saddle point method in our
formulation works best if the initial wave function is a zero
angular momentum state.  Multicentric molecular wave functions
contain many different angular momenta and correspondingly we see
small inaccuracies when we use the saddle point method for
molecules.

In contrast to previous reports on saddle point methods in the velocity
gauge SFA~\cite{Requate03}, we did not find a critical lowest intensity below which the
saddle point method fails.
Even though we find small errors which are direct consequences of
using the saddle point method instead of a numerical evaluation of the
transition amplitude, the error is nearly constant for a wide range of
intensities and is small compared to the large variations in the absolute rates.
Furthermore, we should keep in mind that the SFA itself is only the leading
order of an \textit{S}-matrix series. The small error in the saddle point
evaluation may therefore turn out to be insignificant compared to, e.g.,
neglecting the Coulomb interaction in the final state~\cite{Becker01b}.

We conclude that the saddle point method in the present two-term
version can be used with advantage for long wavelengths and high
intensities when many photon absorptions lead to ionization. In this
case, the transition amplitude is difficult to evaluate numerically
since the integrand oscillates rapidly. The SFA also applies to
non-monochromatic fields, e.g., a few cycle pulse. The transition
amplitude is then calculated by an integral over the full duration
of the pulse. Numerical integration by standard Gaussian quadrature
requires thousands of function evaluations to obtain
convergence~\cite{Martiny06}. It would clearly be desirable to
extend the saddle point method to such a situation where we need
just a few saddle point evaluations.

\begin{acknowledgments}
We thank V. N. Ostrovsky for useful discussions. L.B.M. is supported
by the Danish Natural Science Research Council (Grant No.
21-03-0163) and the Danish Research Agency (Grant. No.
2117-05-0081).
\end{acknowledgments}

\appendix
\section{Complex momentum}
\label{sec:momentum} In connection with Eq.~\eqref{eqn:aqn2}, we
introduced the kinematical momentum
\begin{equation}
  \bm Q(t) = \bm q - \frac{\bm F_0}{\omega} \sin(\omega t).
\end{equation}
We let the laser polarization point in the $z$ direction and find
the squared momentum for a complex time $t = t_r + i t_i$
\begin{equation}
  Q(t)^2 = q_x^2 + q_y^2 + [q_z - \frac{F_0}{\omega} \sin(\omega t)]^2
\end{equation}
with $\sin(\omega t) = \sin(\omega t_r)\cosh(\omega
  t_i) +i \cos(\omega t_r)\sinh(\omega t_i) $.
We wish to calculate $Q(t)$ along the path $-\mathcal{P}_3$ in
Fig.~\ref{fig:path}. The path is parametrized as $\{t = t_r + i
t_i\, |\, 0\le t_r \le T\}$ with the imaginary part $t_i$ fixed.
\begin{figure}
  \begin{center}
    \includegraphics[width=\columnwidth]{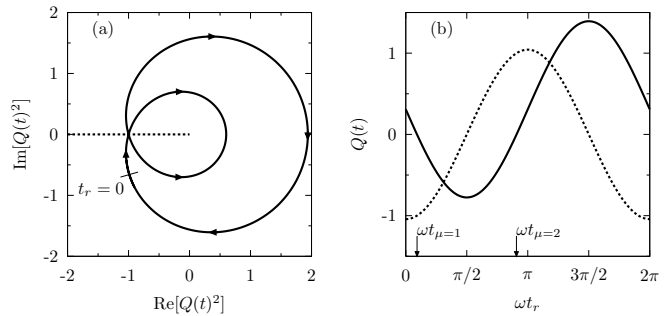}
  \end{center}
  \caption{(a) $Q(t)^2$ in the complex plane along the
  path $-\mathcal{P}_3$ in Fig.~\ref{fig:path}. (b) Real (solid) and
  imaginary (dashed) part of $Q(t)$ according to the
  definition Eq.~\eqref{eqn:qtbranch}.}
  \label{fig:qt}
\end{figure}
In the polar form $Q(t)^2 = |Q(t)^2| e^{i\theta(t)}$, we define the
domain of the phase of $Q(t)^2$ between $-\pi < \theta \le \pi$.
When we calculate the square root, we lie a branch cut along the
negative semi-axis and change the sign when the branch cut is
crossed. In Fig.~\ref{fig:qt}~(a), we show a parametric plot of
$Q(t)^2$ along the path $-\mathcal{P}_3$ from Fig.~\ref{fig:path}.
Using the definition above, we find $Q(t)$
\begin{equation}
  \label{eqn:qtbranch}
  Q(t) = \left\{
  \begin{array}{ll}
   +\sqrt{|Q(t)^2|} e^{i\theta(t)/2} & \textrm{Outer loop} \\
   -\sqrt{|Q(t)^2|} e^{i\theta(t)/2} & \textrm{Inner loop}
  \end{array} \right.,
\end{equation}
where the outer and inner loops refer to Fig.~\ref{fig:qt}~(a). In
Fig.~\ref{fig:qt}~(b), we show the real and imaginary parts of
$Q(t)$ along $-\mathcal{P}_3$. As in Fig.~\ref{fig:path}, we have
$\kappa = 1$ for the ground state of hydrogen. We see that $Q(t) =
-i\kappa$ at the left saddle point ($\omega t_r \approx 0.29$) and
$Q(t) = i\kappa$ at the right saddle point ($\omega t_r \approx
2.85$), which leads to the factor $(\pm 1)^l$ in
Eq.~\eqref{eqn:aqn_saddle}.

\end{document}